\documentclass[aps,prb,floatfix,superscriptaddress,twocolumn,footinbib,longbibliography]{revtex4-2}
\usepackage{amssymb}
\usepackage{amsmath}
\usepackage{amsfonts}
\usepackage{appendix}
\usepackage{bm}
\usepackage[caption=false]{subfig}
\usepackage{graphicx}
\usepackage{epsfig}
\usepackage{epstopdf}
\usepackage{balance}
\usepackage[dvipsnames]{xcolor}
\usepackage{calc}
\usepackage{natbib}
\usepackage[colorlinks,
linkcolor=blue,
anchorcolor=blue,
citecolor=blue,
urlcolor=blue]{hyperref}
\usepackage{lipsum}
\usepackage{braket}
\usepackage{subfig}
\usepackage{hyperref}
\usepackage{verbatim}
\newcommand{\mT}{{\mathcal{T}}}

\begin{document}
\title{Violation of the Wiedemann-Franz law in coupled thermal and power transport of optical waveguide arrays}
\author{Meng Lian}
\affiliation{School of Physics, Institute for Quantum Science and Engineering and Wuhan National High Magnetic Field Center,
Huazhong University of Science and Technology, Wuhan 430074,  China}
\author{Yin-Jie Chen}
\affiliation{School of Physics, Institute for Quantum Science and Engineering and Wuhan National High Magnetic Field Center,
Huazhong University of Science and Technology, Wuhan 430074,  China}
\author{Yue Geng}
\affiliation{School of Physics, Institute for Quantum Science and Engineering and Wuhan National High Magnetic Field Center,
Huazhong University of Science and Technology, Wuhan 430074,  China}
\author{Yuntian Chen}     
\affiliation{School of Optical and Electronic Information, Huazhong University of Science and Technology, Wuhan 430074, China}
\affiliation{Wuhan National Laboratory for Optoelectronics, Huazhong University of Science and Technology, Wuhan 430074, China}
\author{Jing-Tao L{\"u}}     
\email{jtlu@hust.edu.cn}
\affiliation{School of Physics, Institute for Quantum Science and Engineering and Wuhan National High Magnetic Field Center,
Huazhong University of Science and Technology, Wuhan 430074,  China}
\begin{abstract}
In isolated nonlinear optical waveguide arrays with bounded energy spectrum,  simultaneous conservation of energy and power of the optical modes enables study of coupled thermal and particle transport in the negative temperature regime. 
Here, based on exact numerical simulation and rationale from Landauer formalism, we predict generic violation of the Wiedemann-Franz law in such systems. This is rooted in the spectral decoupling of thermal and power current of optical modes, and their different temperature dependence. 
Our work extends the study of coupled thermal and particle transport into unprecedented regimes, not reachable in natural condensed matter and atomic gas systems. 
%
\end{abstract}

\maketitle

Precisely-engineered optical systems have enabled study of problems originated from condensed matter and other branches of physics in clean and controllable setups\cite{Yablonovitch_1987,Sajeev_1987,Haldane_2008}. Recently, such effort has been extended from linear, Hermitian to nonlinear, non-Hermitian systems\cite{miri2019exceptional,el2018non,wright_physics_2022}. An important insight is description of the complex behavior of weakly nonlinear coupled optical waveguide arrays within the framework of statistical thermodynamics\cite{wu_thermodynamic_2019,parto_thermodynamic_2019,makris_statistical_2020}. It provides fresh new insight in understanding puzzling optical phenomena, such as beam self-cleaning \cite{krupa_spatial_2017,liu_kerr_2016,niang_spatial_2019}, 
spatiotemporal solitons\cite{herr_temporal_2014,wright_controllable_2015,kalashnikov_stabilization_2022},
mode-locking \cite{wright_spatiotemporal_2017},  
 and in designing novel devices to manipulate optical waves\cite{wu_thermodynamic_2019}, from the point of view of statistical mechanics
\cite{jung_thermal_2022,pyrialakos_thermalization_2022,shi_controlling_2021,leykam_probing_2021,bloch_non-equilibrium_2022,xiong_k-space_2022,baudin_classical_2020}.

When an isolated system initially deviates from equilibrium, it relaxes to equilibrium through thermal and particle transport mediated by nonlinear interactions within the system.
As a cornerstone result of linear irreversible thermodynamics\cite{callen1951irreversibility}, such coupled transport of thermal and particle current is characterized by the transport matrix. Its diagonal and off-diagonal elements correspond to particle ($G$), thermal ($G_T$) conductance (or conductivity) and thermo-particle (Seebeck and Peltier) transport coefficients, respectively. Study of these transport properties has gained invaluable information on the equilibrium and nonequilibrium properties of condensed matter and atomic gases\cite{brantut_thermoelectric_2013,nietner_transport_2014,gallego-marcos_nonequilibrium_2014,hausler_interaction-assisted_2021,2013arXiv1306.4018H,PhysRevLett.113.170601,rancon_bosonic_2014,uchino_bosonic_2020,husmann_connecting_2015,husmann_breakdown_2018}. The celebrated Wiedemann-Franz (W-F) law links the thermal and particle transport coefficients of the system, with the proportionality constant $L=G_T/(GT)$ being constant in Fermi liquids $L_0=k_B^2\pi^2/3$ with $k_B$ the Boltzmann constant. Break down of the W-F law indicates decoupling of thermal and particle transport\cite{lee_anomalously_2017,husmann_breakdown_2018}, whose origin ranges from strong inter-particle interaction to singular quasi-particle density of states.

\begin{figure}
\centering
  \subfloat{\label{fig:1a}\includegraphics[width=0.45\textwidth]{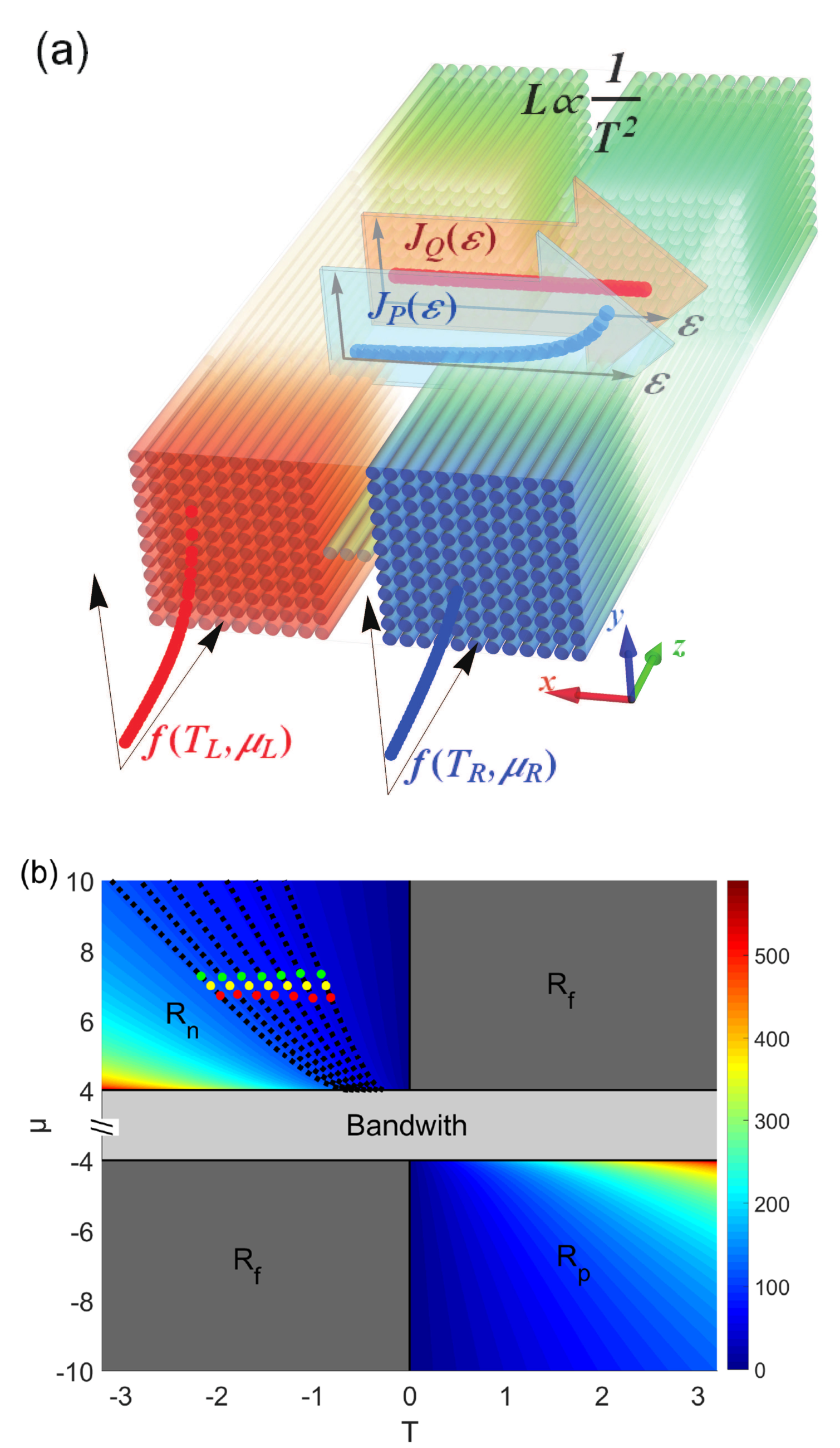}}~~
  \subfloat{\label{fig:1b}\includegraphics[width=0\textwidth]{picture1}}~~\\
\caption{(a) Transport junction made from nonlinear waveguide arrays with square reservoirs. The two reservoirs are connected by one dimensional chain. The onsite energy of nodes is set to zero as the energy reference.
Only nearest neighbor coupling $\kappa_{0}=1$ is considered. The system is initialized at lower end of the waveguides, and it relaxes toward equilibrium during propagating along $z$ direction, through heat and power transport across the junction in the $x$ direction. A generic violation of the Wiedemann-Franz law is predicted with the Lorentz number $L\propto T^{-2}$, indicating decoupled thermal ($J_Q$) and power ($J_P$) current [Eqs.~(\ref{eq:jp}-\ref{eq:jq})]. Shown in the inset  is the energy dependence of $J_Q(\varepsilon)$ and $J_P(\varepsilon)$ for constant transmission $\mathcal{T}$.  
(b) $(T, \mu)$ phase diagram of the square reservoir. $R_{\rm p}$ and $R_{\rm n}$ represent the positive and negative temperature regime, respectively. No equilibrium state exists in the $R_{\rm f}$ regime. Power ($P$) for given $(T, \mu)$ is represented by the colored code. The black dashed lines are the iso-power lines. The red (green) dots are the initial parameters of the left (right) reservoir, while the yellow dots are their average $\bar{T}$ and $\bar{\mu}$.}
\end{figure}

In this work, by numerically tracking the thermalization process of isolated transport junctions made from weakly nonlinear coupled waveguides within the microcanonical ensemble, we study the coupled power\footnote{It plays the role of particle number here. Thus, power conservation corresponds to particle number conservation.}  and thermal transport in such junctions. This is made possible due to simultaneous power and energy conservation. The same condition is difficult to realize in other bosonic systems like phonons or magnons, where there is no particle number conservation. We predict a generic violation of the W-F law in such junctions, with the Lorenz number $L \propto 1/T^2$ (Fig.~\ref{fig:1a}). It does not depend on details of the lattice structures, model parameters, 
 and is valid in both positive and negative temperature regimes. We provide a rationale of this result using the Landauer transport theory in the classical limit.
These results illustrate the opportunity to study fundamental problems of statistical thermodynamics in unprecedented regimes using engineered optical systems \cite{baldovin2021statistical,marques2023observation}. 
{\bf Model --}
We consider transport junction made from one-dimensional waveguide arrays. The junction consists of two large but finite-size reservoirs connected by a chain.  We study coupled heat and power transport properties across the chain. As shown in Fig.~\ref{fig:1a}, each node represents a waveguide, along which ($z$ direction) the optical wave can propagate. In the reservoir the nodes are periodically arranged. This is a prototypical junction structure to study transport properties of solid-state \cite{agrait2003quantum} and cold atom systems\cite{krinner2017two}. 

Propagation of the optical modes along the waveguides is described by 
the discrete nonlinear Schr\"odinger equation (DNSE)\cite{lederer_discrete_2008,agrawal_applications_2001}
\begin{equation} 
i\frac{d\psi _{m}}{dz}+\kappa \sum_{\{n\}}{\psi _{n}}+\chi \left| \psi _{m} \right|^2\psi _{m}=0,
  \label{equ:DNES}
\end{equation}
with dimensionless parameters $\psi_m$, $\kappa$, $\chi$ representing the complex wave amplitude, the nearest neighbour coupling between waveguides and Kerr-type nonlinear coefficient, respectively.  
Here the coordinate $z$ along waveguide plays the role of time in standard Schr\"odinger equation, and $\{n\}$ includes all the nearest neighbours of $m$. 

For weak nonlinear interactions, we can  diagonalize the linear term and obtain the corresponding eigenmode (supermode) propagating constant $\beta_k$ and corresponding vector $\varphi_k$. For a given state $\psi$, the modal occupancy is obtained by projection onto each supermode 
$\left|c_k\right|^2=\left| \langle \varphi _k | \psi \rangle \right|^2$.
The system evolves with conserved internal energy
$U=\sum_{k=1}^M{\varepsilon _k\left| c_k \right|^2}	$
and optical power
$P=\sum_{k=1}^M{\left| c_k \right|^2}$,
with the total number of modes $M$, the eigen energy defined by the negative of $\beta$ as $\varepsilon_k = -\beta_k$, and $\beta _1\ge \beta _2\ge \cdots \ge \beta _M$. The system thus has a bounded energy spectrum between $[\varepsilon_1, \varepsilon_M]$.
The weak nonlinear part introduces coupling among different supermodes, such that the system can thermalize to an equilibrium state through energy and power redistribution among supermodes. It plays a role similar to molecular collisions in the ideal gas model.

{\bf Equilibrium thermodynamic theory --} 
We can use a recently developed optical thermodynamic theory to describe each reservoir\cite{wu_thermodynamic_2019}. The system internal energy can be written as (omitting the reservoir index)
\begin{align}
	U = M T + \mu P,
 \label{eq:eos}
\end{align}
where $T$ is the dimensionless optical temperature, $\mu$ is the chemical potential. An optical entropy can be defined as 
\begin{align}
S=\sum_{k=1}^M{\ln \left| c_k \right|^2}. 
\end{align}
Using the maximum entropy principle, at thermal equilibrium, the mode occupancy follows the classical Rayleigh-Jeans (R-J) distribution 
\begin{align}
	f(\varepsilon_k) = \left| c_k \right|^2=\frac{T}{\varepsilon _k-\mu}.
	\label{eq:rj}
\end{align}

\begin{figure}
\centering
  \subfloat{\label{fig:3a}\includegraphics[width=0.48\textwidth]{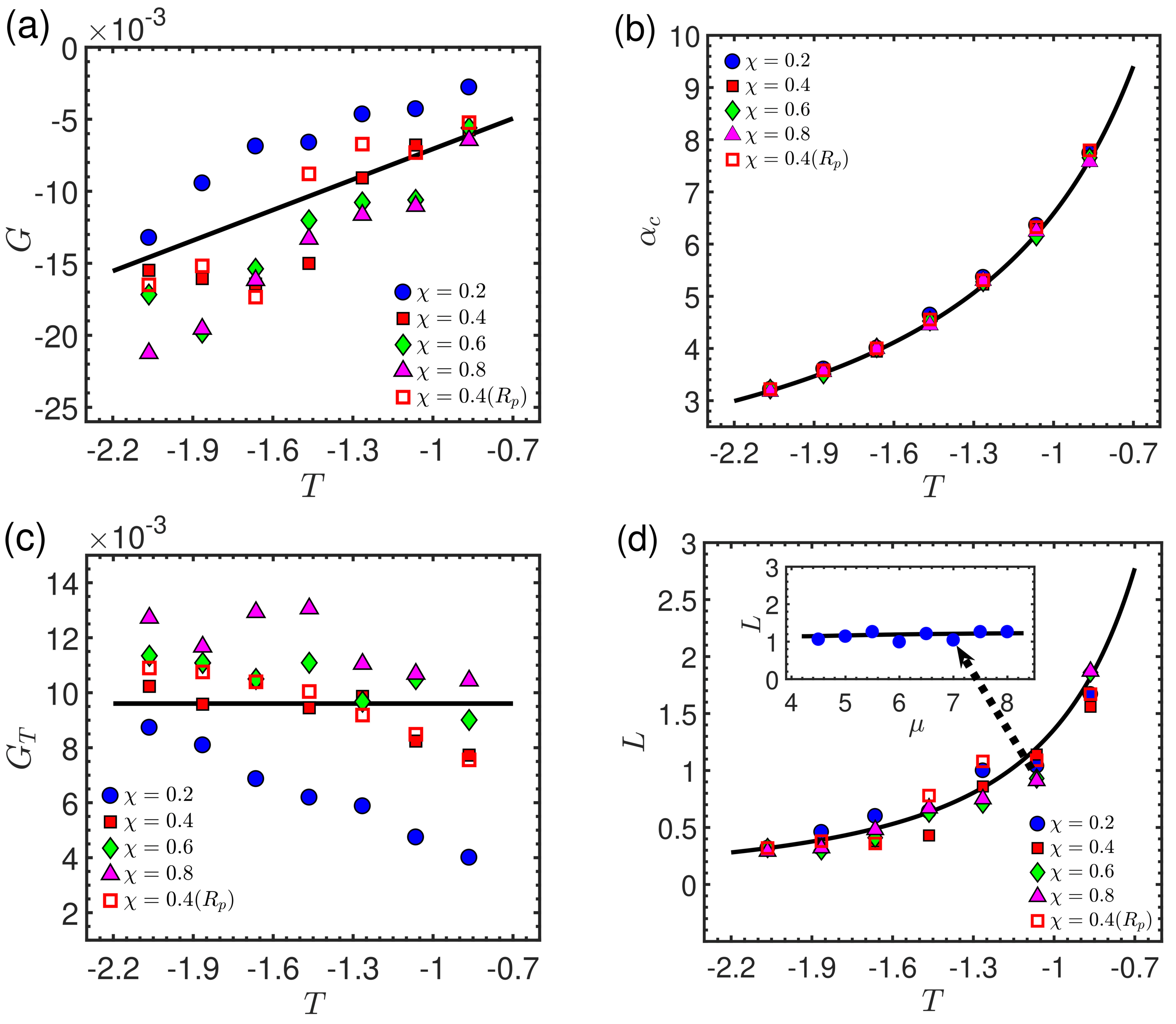}}~~
  \subfloat{\label{fig:3b}\includegraphics[width=0\textwidth]{picture2}}~~\\
  \subfloat{\label{fig:3c}\includegraphics[width=0\textwidth]{picture2.eps}}~~
  \subfloat{\label{fig:3d}\includegraphics[width=0\textwidth]{picture2}}~~
\caption{Dependence of transport coefficients on temperature ($T$) for different nonlinear strengths $\chi$ at $\bar{\mu}=7$. (a) Conductance $G$,  (b)Seebeck coefficient $\alpha _c$, (c) Thermal conductance $G_T$, (d) Lorenz number $L$. The inset shows dependence of $L$ on chemical potential $\mu$ at $T = -1.06$ and $\chi = 0.2$. The black solid lines are theoretical results from the Landauer formalism. Results at positive temperature with $\chi=0.4$ are shown with empty squares. The sign of $T$ and $G$ at positive $T$ are reversed to show in the figure. They are very close to those at corresponding negative temperature (filled squares). }
     \label{fig:3}
\end{figure}

One prominent feature of the system is that, for given power $P$, the 
optical entropy $S$ no longer varies monotonically with the internal energy $U$ due to bounded energy spectrum. Consequently, the system can reach the  negative optical temperature regime\cite{wu_thermodynamic_2019,wu_entropic_2020}.
From Eqs.~(\ref{eq:eos}-\ref{eq:rj}), we have plotted the phase diagram of the reservoir with square lattice. As shown in Fig.~\ref{fig:1b}, only $\left\{ \left( T,\mu \right) |T<0,\mu >\varepsilon _M \right\}$ and $\left\{ \left( T,\mu \right) |T>0,\mu <\varepsilon _1 \right\}$ can be visited, and the other regimes are forbidden($R_f$). This follows from the requirement $\left| c_k \right|^2 \ge 0$.  Moreover, the chemical potential is out of the band spectrum of the system, i.e., $\mu$ is below (above) the lowest (highest) energy of the spectrum for positive (negative) temperature. The positive and negative temperature regimes are anti-symmetric with each other in the phase diagram when neglecting the nonlinear effect.  
In the following, we focus on the negative temperature regime and numerically check that this anti-symmetry is still approximately valid in the weakly nonlinear regime (Fig.~\ref{fig:3}). 

{\bf Transport coefficients in microcanonical ensemble --} 
Transport theory is normally formulated with open boundaries using grand canonical ensemble.  Due to the finite size of present junction, it is convenient to utilize the microcanonical setup. We extract the transport coefficients by following the thermalization process of a microcanonical system from given initial conditions.
Using the theory of irreversible thermodynamics, 
we can obtain the evolution equations of $\Delta P$ and $\Delta T$ (See Appendix~\ref{C}). 
For symmetric junctions with identical left and right reservoirs, 
they read
\begin{equation} 
  \tau _0\frac{\partial}{\partial z}\left( \begin{array}{c}
    \Delta P\\
    \Delta T\\
  \end{array} \right) =-\left( \begin{matrix}
    1&		-\kappa \alpha\\
    -\frac{\alpha}{\ell \kappa}&		\frac{L+\alpha ^2}{\ell}\\
  \end{matrix} \right) \left( \begin{array}{c}
    \Delta P\\
    \Delta T\\
  \end{array} \right) 
  \label{equ:5}
\end{equation}
where we have defined a transport timescale $\tau _0=\kappa /\left( 2G \right)$ .
Here, the reservoir properties $\kappa =\left. \left( \partial P/\partial \mu \right) \right|_T$, $\alpha _r=-\left. \left( \partial \mu /\partial T \right) \right|_P$, $C_P=\left. T\left( \partial S/\partial T \right) \right|_P$ are the compression coefficient, Seebeck coefficient and the heat capacity at constant power, respectively,
with $l=C_P/T\kappa$.
They can be obtained from the thermodynamic relations [Eqs.~(\ref{equ:kappa}-\ref{equ:CP})]. 
The effective Seebeck coefficient $\alpha=\alpha_r-\alpha_c$ characterizes the difference between the reservoir itself $\alpha_r$ and the whole junction $\alpha_c$.

\begin{figure}
\centering
\includegraphics[width=0.48\textwidth]{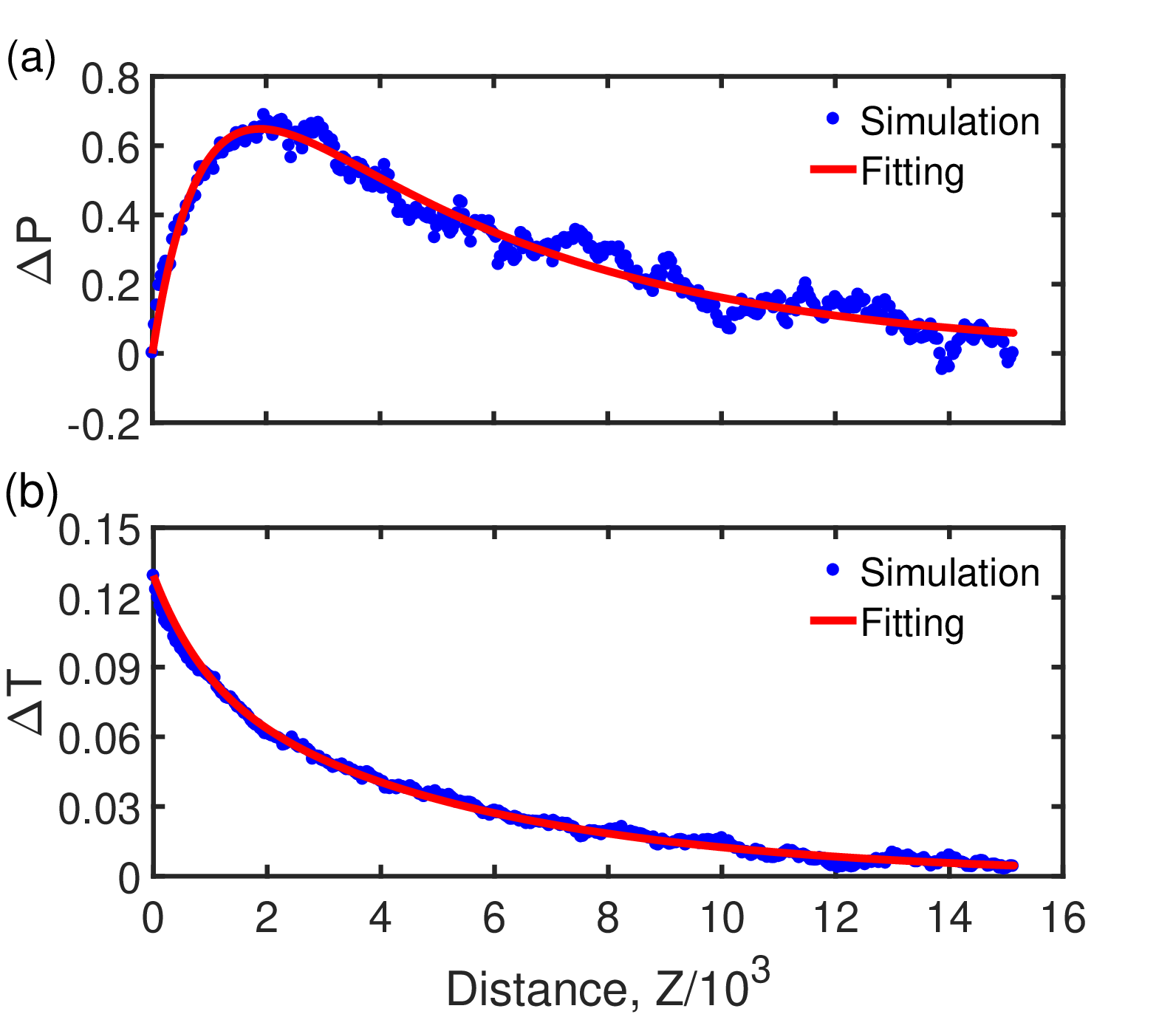}
\caption{\label{fig:2}Evolution of (a) power ($\Delta P$) and (b) temperature ($\Delta T$) deviations from the average values $\bar{T}=-1.065$ and $\bar{\mu}=7$, with $\Delta T=0.13$, $\Delta \mu=-0.7$, $\chi =0.2$. The blue dots are numerical results, and the red line is fitting from Eqs.~(\ref{equ:6}-\ref{equ:7}).}
\end{figure}

The general solution of the above equations are
\begin{widetext}
	\begin{align}
		\Delta P\left( z \right) =\left( \frac{\Omega +\delta -1}{2\Omega}\Delta P_0+\frac{\alpha \kappa}{2\Omega}\Delta T_0 \right) e^{-\frac{z}{\tau _>}}\left[ 1+\frac{\left( \Omega -\delta +1 \right) \Delta P_0-\alpha \kappa \Delta T_0}{\left( \Omega +\delta -1 \right) \Delta P_0+\alpha \kappa \Delta T_0}e^{-\frac{z}{\tau _<}} \right],
  \label{equ:6}\\
  \Delta T\left( z \right) =\left( \frac{\Omega -\delta +1}{2\Omega}\Delta T_0+\frac{\alpha}{2l\kappa \Omega}\Delta P_0 \right) e^{-\frac{z}{\tau _>}}\left[ 1+\frac{l\kappa \left( \Omega +\delta -1 \right) \Delta T_0-\alpha \Delta P_0}{l\kappa \left( \Omega -\delta +1 \right) \Delta T_0+\alpha \Delta P_0}e^{-\frac{z}{\tau _<}} \right]
  \label{equ:7}
	\end{align}
\end{widetext}
where $\delta =\left[ 1+\left( L+\alpha ^2 \right) /l \right] /2$ and $\varOmega =\sqrt{\delta ^2-L/l}$ form the eigenvalues of the evolutionary matrix $\lambda _{\pm}=\delta \pm \varOmega $. 
The fast and slow timescales $\tau_<=\tau _0/2\varOmega$ and $\tau_>=\tau _0/\left( \delta -\varOmega \right)$ indicate the evolution of the system in two stages (Fig.~\ref{fig:2}): (1) A saturation process is characterized by the time scale $\tau_<$, which dominates the initial short time period. This process leads to an initial increase in the absolute value of the particle number deviation until it reaches a maximum . (2) 
A decay process is characterized by the time scale $\tau_>$, which dominates the longer period of time after saturation. 
The three transport coefficients $G$, $\alpha _c$ and $L$ (or $G_T$) can then be acquired by fitting the numerical results . 
Figure~\ref{fig:2} shows results of fitting the transient evolution of the rectangular junction. The details can be found in Appendix~\ref{D}.

{\bf Numerical Results --} 
Figure~\ref{fig:3a}-\ref{fig:3d} shows temperature dependence of the transport coefficients for a symmetric junction with square-lattice reservoirs. 
We notice that, in the negative temperature regime, conductance $G<0$, indicating that the power is transferred from the low to the high chemical potential reservoir when $\varDelta T=0$. This seemingly surprising result does not violate the second law of thermodynamics.
Considering two subsystems that can exchange particles with each other under constant temperature, the second law of thermodynamics requires that the entropy does not decrease during the particle exchange
\begin{equation} 
  dS=\left( \frac{\mu _L}{T_L}-\frac{\mu _R}{T_R} \right) dP \ge 0
  \label{equ:9}
\end{equation}
where $dP$ is the power lost from the left reservoir or gained from the right reservoir.
When $T_L=T_R>0$, the power is transferred from high to low chemical potential. However, when $T_L=T_R<0$, the direction is reversed, meaning $G<0$. Since heat is transferred from high to low temperature in both positive and negative temperature regions, we have $G_T>0$. 

Figure~\ref{fig:3a} shows an approximated linear $T$ dependence of $G$ for all the nonlinear parameters considered. This can be understood qualitatively from the Landauer formalism (Appendix~\ref{sec:lt}). 
%
%
The conductance can be approximated as 
\begin{equation} 
  G=\int {\frac{d\varepsilon}{2\pi}{\mathcal{T}}(\varepsilon) \frac{T}{\left( \varepsilon -\mu \right) ^2}},
  \label{equ:G}
\end{equation}
indicating $G \propto T$ for temperature independent transmission $\mathcal{T}(\varepsilon)$. Moreover, the optical modes contribute to the transport with a weighting factor inversely proportional to the square of the energy deviation from the chemical potential.  

Temperature dependence of the Seebeck coefficient $\alpha_c$ is shown in Fig.~\ref{fig:3b}. It can be well fitted by an inverse $T$ dependence $\alpha_c \propto -1/T$ predicted by the Landauer model. 
%
The little effect of nonlinear interaction on $\alpha_c$ can also be captured by the Landuer result. In fact, $\alpha_c$ can be written as $\alpha _c = K_1/(K_0T)$. $K_0$ and $K_1$ have similar dependence on the nonlinear interaction, they cancel with each other in $\alpha_c$. 

Similar to conductance, the thermal conductance $G_T$ also increases with nonlinearity (Fig.~\ref{fig:3c}). From the Landauer formalism, we get
\begin{equation} 
  G_T=\int {\frac{d\varepsilon}{2\pi}}\mathcal{T}(\varepsilon) -T\alpha _{c}^{2}G.
  \label{equ:15}
\end{equation}
This gives a temperature independent $G_T$. Similar to $G$, the temperature dependence in the numerical results is due to nonlinear interaction.  
The second term on the right side of Eq.~(\ref{equ:15}) represents correction to the thermal conductance due to Seebeck coefficient. Here, its magnitude is comparable to the first term. This is in contrast to the case of electrons in condensed matter system, where the thermoelectric correction is often negligible.  

The variation of $L$ with temperature and chemical potential is shown in Fig.~\ref{fig:3d}, and is once again captured by Landauer's theory. 
It can be found that the Lorenz number is more sensitive to temperature than the chemical potential (Inset).
In the Landauer picture, $G_T \propto T^0$ together with $G \propto T$ leads to $L \propto 1/T^2$, which violates the W-F law. The decrease in $L$ with decreasing temperature comes from the increase in $\left| G \right|$.
Similar to $\alpha _{c}$, we also observe a rather weak dependence on the strength of nonlinear interaction. Thus, the violation of W-F law here is not due to nonlinear interaction. Rather, its origin is similar to that in coherent mescoscopic conductors, which can be fully accounted by the single particle Landauer formalism.


\begin{figure}
	\centering
  \includegraphics[width=0.48\textwidth]{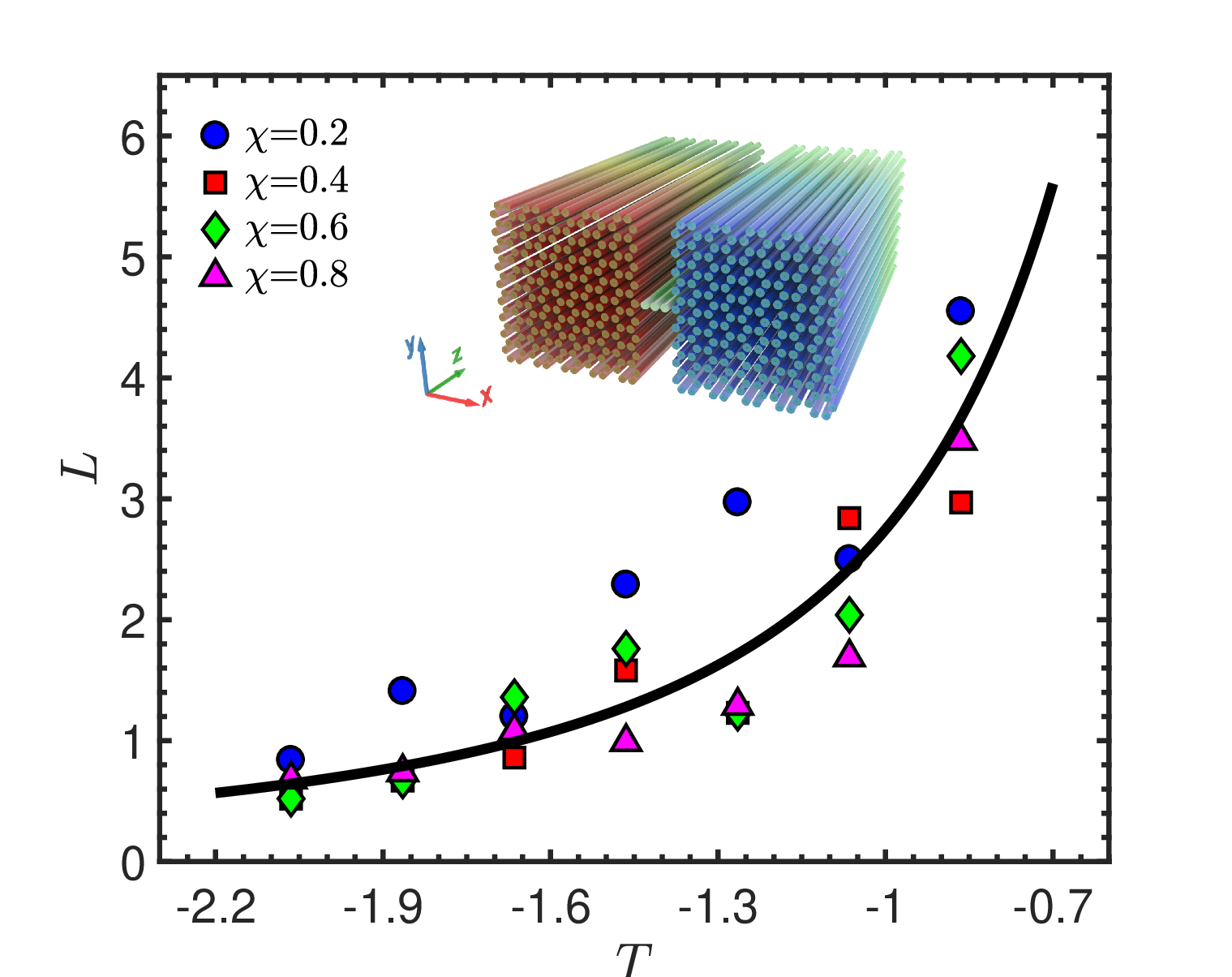}
	\caption{\label{fig:5}Variation of Lorenz number with temperature for different nonlinear strengths at $\bar{\mu}=7$ for a junction with hexagonal honeycomb reserviors (Inset). The black solid line is theoretical result from the Landauer formalism.}
\end{figure}

To further illustrate the violation of the W-F law, we considered a symmetric junction made from hexagonal honeycomb reservoirs with the same bandwidth (Fig.~\ref{fig:1a}).
Junction with hexagonal reservoirs has a broader transmission spectrum compared to that with square reservoirs (Fig.~\ref{fig:T}).
This is due to the two van Hove singularities unique to hexagonal lattice, particularly evident in the DOS (Fig.~\ref{fig:DOS}), giving higher transmission to the modes near the band edges.
This makes the thermal conductivity increase to about twice that of the square case.
As a result, we observe a more severe violation of the W-F law (Fig.~\ref{fig:5}), albeit with the same $1/T^2$ dependence.

{\bf Conclusions --}
We have developed a theory based on microcanonical ensemble 
to extract the transport coefficients of the coupled thermal and power transport in coupled waveguide arrays. We found a systematic violation of the Wiedemann-Franz law with the Lorenz number $L \propto 1/T^2$. 
The numerical results can be understood  from the classical limit of Landauer formula. Our theory based on microcanonical ensemble is compatible to experimental realizations.   
Thus, it paves the way of studying fundamental nonequilibrium thermodynamic processes  in parameters regimes that are unreachable in condensed matter and atomic gas systems. 


This work was supported by the National Natural Science Foundation of China (Grant Nos. 22273029 and 21873033).

\appendix
\renewcommand\thefigure{\thesection\arabic{figure}}  

\section{Evolution equations for $\Delta P$ and $\Delta T$} \label{C}
\setcounter{figure}{0} 
Using the theory of irreversible thermodynamics, we can make connection between the driving affinities $\varDelta \mu$, $\varDelta T$ with the corresponding power ($I_P$) and entropy ($I_S$) fluxes, which follow the Onsager symmetry
\begin{equation} 
\left( \begin{array}{c}
	I_P\\
	I_S\\
\end{array} \right) =\left( \begin{matrix}
	K_0&		K_1/T\\
	K_1/T&		K_2/T^2\\
\end{matrix} \right) \left( \begin{array}{c}
	\Delta \mu\\
	\Delta T\\
\end{array} \right). 
\label{equ:1}
\end{equation}
Positive direction of the current is chosen as from $L$ to $R$ reservoir.
The matrix elements $K_0$, $K_1$ and $K_2$ depend on the state of the system and can be expressed in terms of the power conductance $G$, the thermal conductance $G_T$, the Seebeck coefficient $\alpha _{c}$ and the Lorenz number $L=G_T/TG$,
\begin{equation} 
  \frac{\partial}{\partial z}\left( \begin{array}{c}
    \Delta P\\
    \Delta S\\
  \end{array} \right) =-2G\left( \begin{matrix}
    1&		\alpha _{{c}}\\
    \alpha _{{c}}&		L+\alpha _{{c}}^{2}\\
  \end{matrix} \right) \left( \begin{array}{c}
    \Delta \mu\\
    \Delta T\\
  \end{array} \right) 
  \label{equ:2}
\end{equation}
%
%
%

Due to the weak link between the two reservoirs, bottleneck of the thermalization process is at the central chain. 
For the two reservoirs, the thermalization can be considered as a quasi-static process, where the two reservoirs can reach local equilibrium much faster than the whole junction. This can be verified by tracking the evolution of the system, and  
allows us to use the thermodynamic equation of the reservoirs.
Using the Maxwell's relations $\left. \left( \partial S/\partial \mu \right) \right|_T=\left. \left( \partial P/\partial T \right) \right|_{\mu}$ and $-\left. \left( \partial S/\partial P \right) \right|_T=\left. \left( \partial \mu /\partial T \right) \right|_P$, we can obtain
\begin{eqnarray}
  dP=\left. \frac{\partial P}{\partial \mu} \right|_Td\mu +\left. \frac{\partial P}{\partial T} \right|_{\mu}dT=\kappa d\mu +\alpha _r\kappa dT,
  \label{equ:3}
  \\
  dS=\left. \frac{\partial S}{\partial P} \right|_TdP+\left. \frac{\partial S}{\partial T} \right|_PdT=\alpha _rdP+\kappa ldT
  \label{equ:4}
\end{eqnarray}
where $\kappa =\left. \left( \partial P/\partial \mu \right) \right|_T$, $\alpha _r=-\left. \left( \partial \mu /\partial T \right) \right|_P$, $l=C_P/T\kappa $ are the compression coefficient, Seebeck coefficient and Lorenz number of the reservoir, respectively,
with $C_P=\left. T\left( \partial S/\partial T \right) \right|_P$ the reservoir heat capacity at constant power.  
In the linear response regime, they can be obtained from the equilibrium thermodynamic state at $\bar{T}$ and $\bar{\mu}$ as 
\begin{equation} 
  \kappa =\left. \sum_i^M{\frac{T}{\left( \varepsilon _i-\mu \right) ^2}} \right|_T,
  \label{equ:kappa}
\end{equation}
\begin{equation} 
  \alpha _r=\frac{1}{\kappa}\left. \sum_{i=1}^M{\frac{1}{\varepsilon _i-\mu}} \right|_{\mu},
  \label{equ:alpha}
\end{equation}
\begin{equation} 
  C_P=M-T\alpha _{r}^{2}\kappa.
  \label{equ:CP}
\end{equation}

%

According to the thermodynamic equations, Eqs.~(\ref{equ:3}-\ref{equ:4}), the optical power $P(z)$ and entropy $S(z)$ of the reservoir at position $z$ can be written as
\begin{align} 
  P\left( z \right) &=\kappa \mu \left( z \right) +\alpha _r\kappa T\left( z \right) +A,
  \label{AP_B_1}\\
  S\left( z \right) &=\alpha _rP\left( z \right) +\kappa lT\left( z \right) +B
  \label{AP_B_2}
\end{align}
where $A$ and $B$ are constants related to the initial conditions. We have dropped the reservoir index. When the left and right reservoirs are identical, their thermodynamic parameters are also identical, $A_L=A_R$ and $B_L=B_R$.
Then, the optical power ($\Delta P\left( z \right) $) and entropy ($\Delta S\left( z \right) $) difference can be written as
\begin{align} 
  \Delta P\left( z \right)&=\kappa \Delta \mu \left( z \right) +\alpha _r\kappa \Delta T\left( z \right), 
  \label{AP_B_3}\\
  \Delta S\left( z \right)&=\alpha _r\Delta P\left( z \right)+\kappa l\Delta T\left( z \right).
  \label{AP_B_4}
\end{align}
Using $\Delta P$ and $\Delta T$ to eliminate $\Delta S$ and $\Delta \mu$ in Eq.~(\ref{equ:2}), the evolution equations are obtained
\begin{equation} 
  \tau _0\frac{\partial}{\partial z}\left( \begin{array}{c}
    \Delta P\\
    \Delta T\\
  \end{array} \right) =-\left( \begin{matrix}
    1&		-\kappa \alpha\\
    -\frac{\alpha}{\kappa l}&		\frac{\alpha ^2+L}{l}\\
  \end{matrix} \right) \left( \begin{array}{c}
    \Delta P\\
    \Delta T\\
  \end{array} \right) \equiv -\varLambda \left( \begin{array}{c}
    \Delta P\\
    \Delta T\\
  \end{array} \right) ,
  \label{AP_B_5}
\end{equation}
where we have defined a transport timescale $\tau _0=\kappa /\left( 2G \right)$ and an effective Seebeck coefficient $\alpha =\alpha _r-\alpha _c$. 
The eigen values of matrix $\varLambda $ are 
\begin{equation} 
  \lambda _{\pm}=\frac{1}{2}\left( 1+\frac{\alpha ^2+L}{l} \right) \pm \sqrt{\left( \frac{1}{2}+\frac{\alpha ^2+L}{2l} \right) ^2-\frac{L}{l}}=\delta \pm \varOmega .
  \label{AP_B_6}
\end{equation}
General solutions of Eq.~(\ref{AP_B_5}) can be expressed using the eigen values
\begin{align} 
  \Delta P\left( z \right) &=C_1e^{-\frac{\lambda _+}{\tau _0}z}+C_2e^{-\frac{\lambda _-}{\tau _0}z},
  \label{AP_B_7}\\
  \Delta T\left( z \right) &=C_3e^{-\frac{\lambda _+}{\tau _0}z}+C_4e^{-\frac{\lambda _-}{\tau _0}z},
  \label{AP_B_8}
\end{align}
where the coefficients $C_1$, $C_2$, $C_3$ and $C_4$ are determined by the initial conditions.
Assuming that the initial power and temperature difference are $\Delta P_0$ and $\Delta T_0$, from Eqs.~(\ref{AP_B_5}),  (\ref{AP_B_7}) and  (\ref{AP_B_8}) we get
\begin{align} 
  C_1&=\frac{1+\Omega -\delta}{2\Omega}\Delta P_0-\frac{\kappa \alpha}{2\Omega}\Delta T_0,
  \label{AP_B_9}\\
  C_2&=\frac{\Omega +\delta -1}{2\Omega}\Delta P_0+\frac{\kappa \alpha}{2\Omega}\Delta T_0,
  \label{AP_B_10}\\
  C_3&=-\frac{\alpha}{2\kappa l\Omega}\Delta P_0+\frac{\delta +\Omega -1}{2\Omega}\Delta T_0,
  \label{AP_B_11}\\
  C_4&=\frac{\alpha}{2\kappa l\Omega}\Delta P_0+\frac{\Omega -\delta +1}{2\Omega}\Delta T_0.
  \label{AP_B_12}
\end{align}
Substituting into Eqs.~(\ref{AP_B_7}) and  (\ref{AP_B_8}), we can obtain the evolution equations for $\Delta P$ and $\Delta T$.

\subsection{Numerical simulation}
\setcounter{figure}{0} 
We choose the initial condition of each reservoir by fixing its optical temperature $T_\alpha$ and chemical potential $\mu_\alpha$.  
The corresponding optical power and internal energy can then be obtained. We perform calculations in the linear response regime with $T_L-T_R=\Delta T\ll \bar{T}$, $\mu _L-\mu _R=\Delta \mu \ll \bar{\mu}$. When the left and right reservoirs are identical, the average temperature and chemical potential are $\bar{T}=\left( T_L+T_R \right) /2$ and $\bar{\mu}=\left( \mu _L+\mu _R \right) /2$, respectively. 
Figure~\ref{fig:1b} shows seven sets of parameters at fixed chemical potential, all selected on iso-power lines while ensuring the small temperature and chemical potential difference.
The fourth-order Runge-Kutta method was used to solve Eq.~(\ref{equ:DNES}).  
The final result for each given initial condition is ensemble averaged by choosing different random phases for $c_{i}$. 
%
%
%
Fig.~\ref{fig:2} shows the calculated results for a symmetric junction with square-lattice reservoirs at $\bar{T}=-1.065$ and $\bar{\mu}=7$.
At the initial stage, driven by temperature and chemical potential deviations, the optical power is rapidly transferred from the cold reservoir to the hot reservoir. After $\Delta P$ reaches saturation, the system is brought to equilibrium by a slow relaxation process.

\subsection{Data Fitting} \label{D}
\begin{figure}
	\centering
	\subfloat{\label{fig:2a}\includegraphics[width=0.38\textwidth]{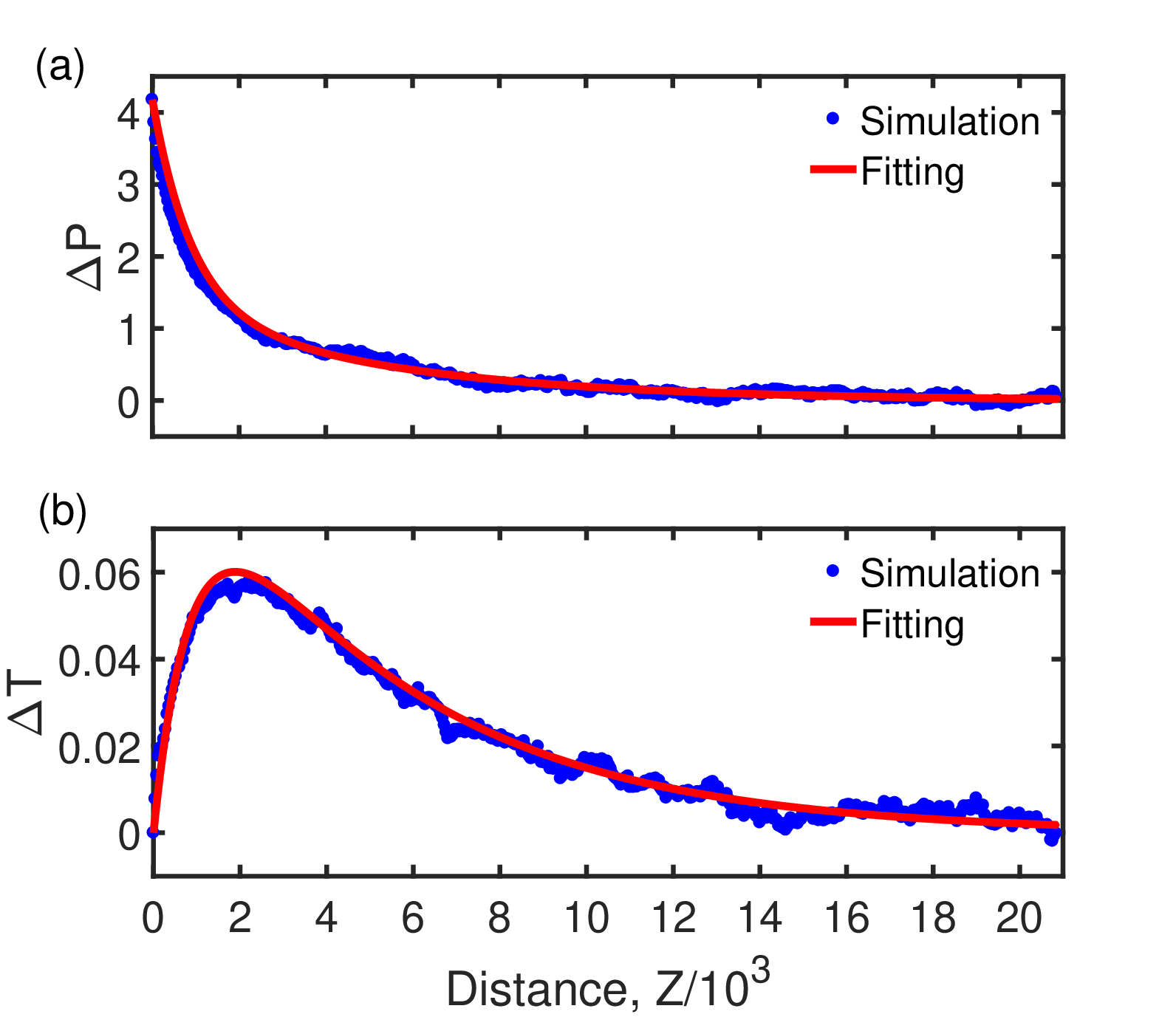}}~~\\
	\subfloat{\label{fig:2b}\includegraphics[width=0.38\textwidth]{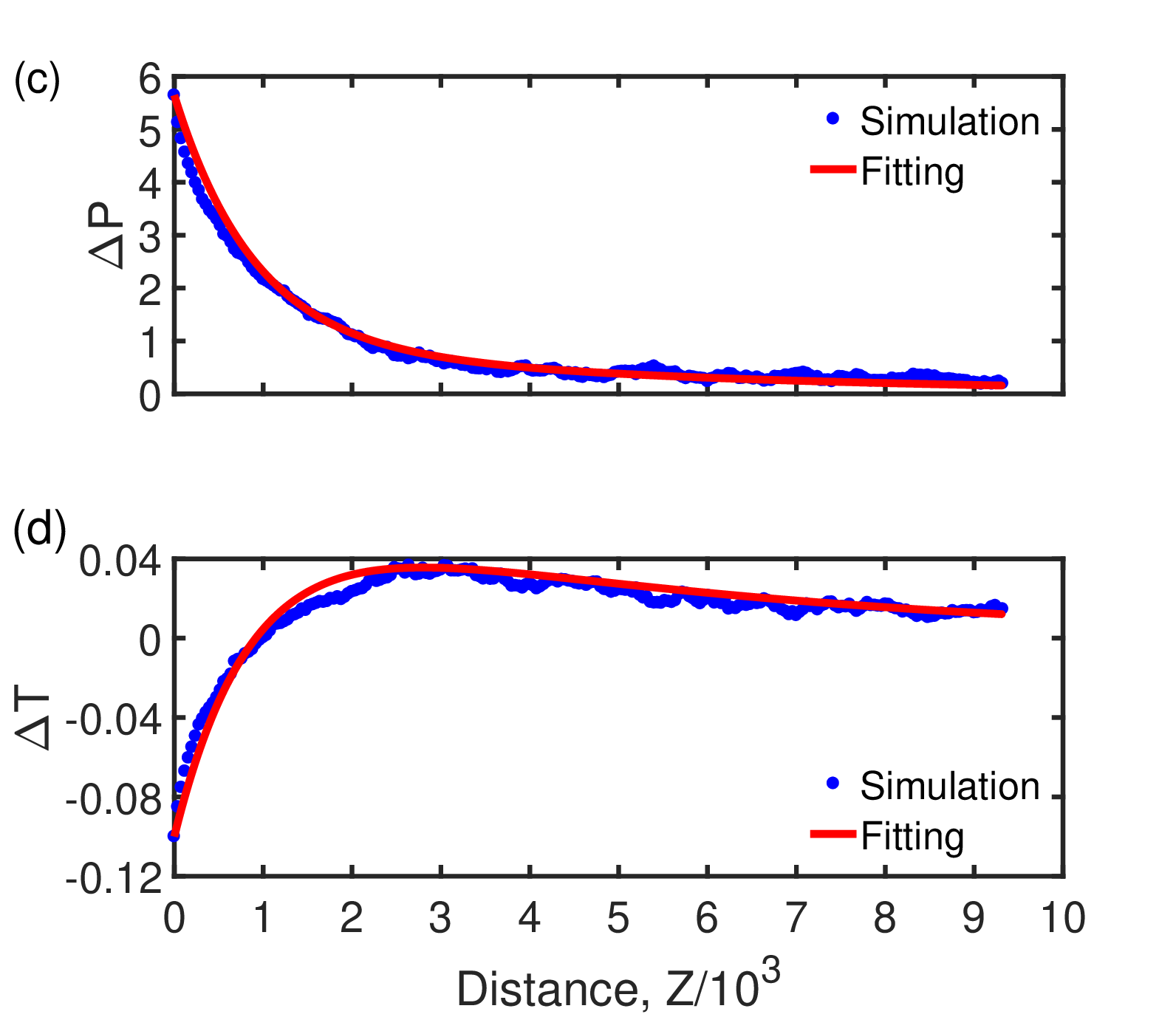}}~~
	\caption{\label{fig:test} Evolution of power ($\varDelta P$) temperature ($\varDelta T$) difference as a function of propagating distance $z$ for two sets of parameters. (a-b) $\Delta P_0=4.1787$, $\Delta T_0=0$. (c-d) $\Delta P_0=5.6458$, $\Delta T_0=-0.1$. The fitting is performed using the same set of transport coefficients in (a-b) and (c-d).}
\end{figure}

\begin{figure}
	\centering
	\subfloat{\label{fig:test_S}\includegraphics[width=0.46\textwidth]{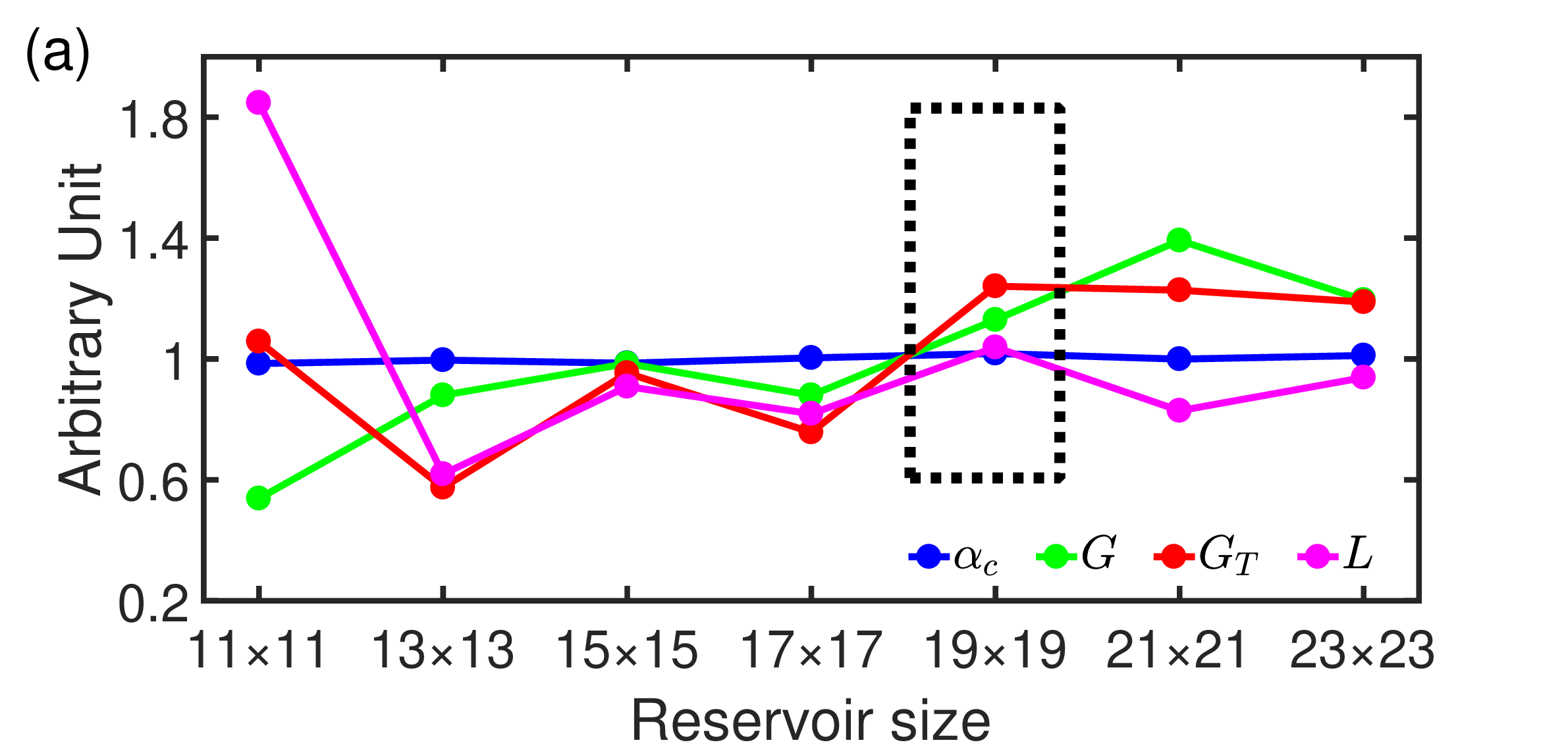}}~~\\
	\subfloat{\label{fig:test_H}\includegraphics[width=0.46\textwidth]{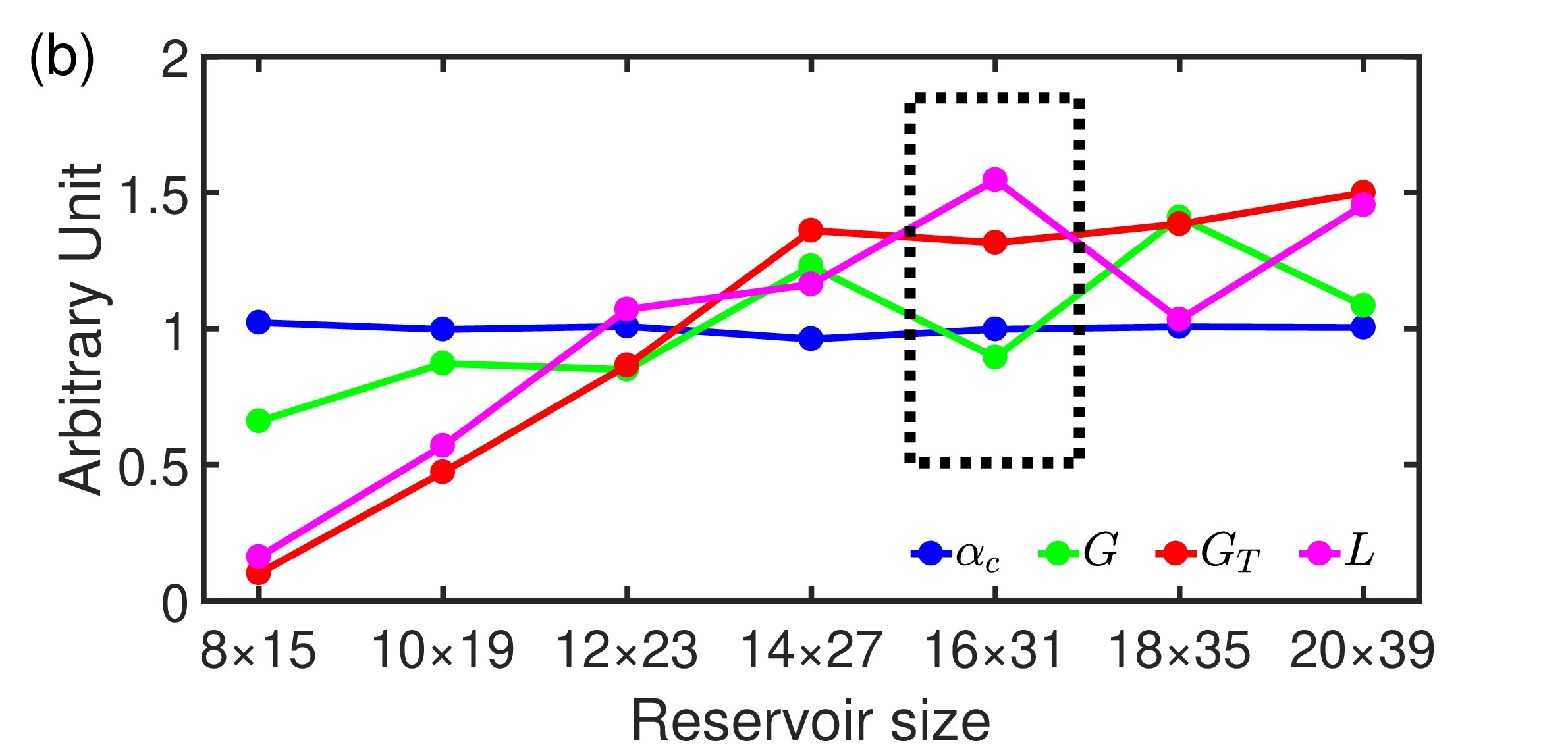}}~~
	\caption{Convergence tests for two types of reservoirs with square (a) and hexagonal honeycomb (b) lattices. All transport coefficients are scaled to the same order of magnitude. The dashed boxes encloses reservoir size used in the main text. } 
\end{figure}
We use Eqs.~(\ref{equ:6}) and (\ref{equ:7}) to fit the evolution of $\varDelta P$ and $\varDelta T$ simultaneously by scanning the chain parameter space so that the residual $R$ is minimized, which is defined as\cite{hausler_interaction-assisted_2021}
\begin{equation} 
  R=\sum_i{\left[ \left( \frac{\Delta P\left( z_i \right) -\Delta P_i}{\bar{P}} \right) ^2+\left( \frac{\Delta T\left( z_i \right) -\Delta T_i}{\bar{T}} \right) ^2 \right]}.
  \label{equ:RES}
\end{equation}
Here, $\Delta P\left( z_i \right)$ and $\Delta T\left( z_i \right)$ are obtained from Eqs.~(\ref{equ:6}) and (\ref{equ:7}) at position $z_i$, $\Delta P_i$ and $\Delta T_i$ are the numerically calculated results from Eq.~(\ref{equ:DNES}) at $z_i$, and $\bar{P}$ and $\bar{T}$ are averages of $\Delta P_i$ and $\Delta T_i$.
Since the calculation is in the linear response regime, our results are only related to the equilibrium temperature and chemical potential of the system, independent of the initial biases $\Delta P_0$ and $\Delta T_0$. This is checked in Fig.~\ref{fig:test} where evolution from two sets of different initial conditions is fitted by the same set of transport coefficients. 

\subsection{Convergence tests} \label{B}
We performed convergence tests for symmetric junctions formed by square and hexagonal honeycomb lattices, respectively, to ensure that the fitted transport coefficients do not depend on the size of the reservoirs. 
The results are summarized in Fig.~(\ref{fig:test}).
Parameters used in the main text are enclosed by dotted boxes.

\section{Landauer transport theory}
\label{sec:lt}
\setcounter{figure}{0} 
The DNSE corresponds to an effective tight-binding Hamiltonian for the junction\cite{wu_thermodynamic_2019,rasmussen_statistical_2000}
\begin{equation} 
  H\left( \psi _m,i\psi _{m}^{*} \right) =\sum_{\{m,n\}}{\kappa \psi _{m}^{*}\psi _n}+\frac{1}{2}\chi \sum_m {\left| \psi _m \right|^4}.
      \label{equ:H}
\end{equation}
Here, $\psi _{m}$ and $i\psi _{m}^{*}$ form the canonical conjugate pairs,  $\{m,n\}$ are nearest neighbour pairs. The first and the second term corresponds to the linear and nonlinear part of the Hamiltonian.  

\begin{figure}[h]
\centering\subfloat{\label{fig:T}\includegraphics[width=0.24\textwidth]{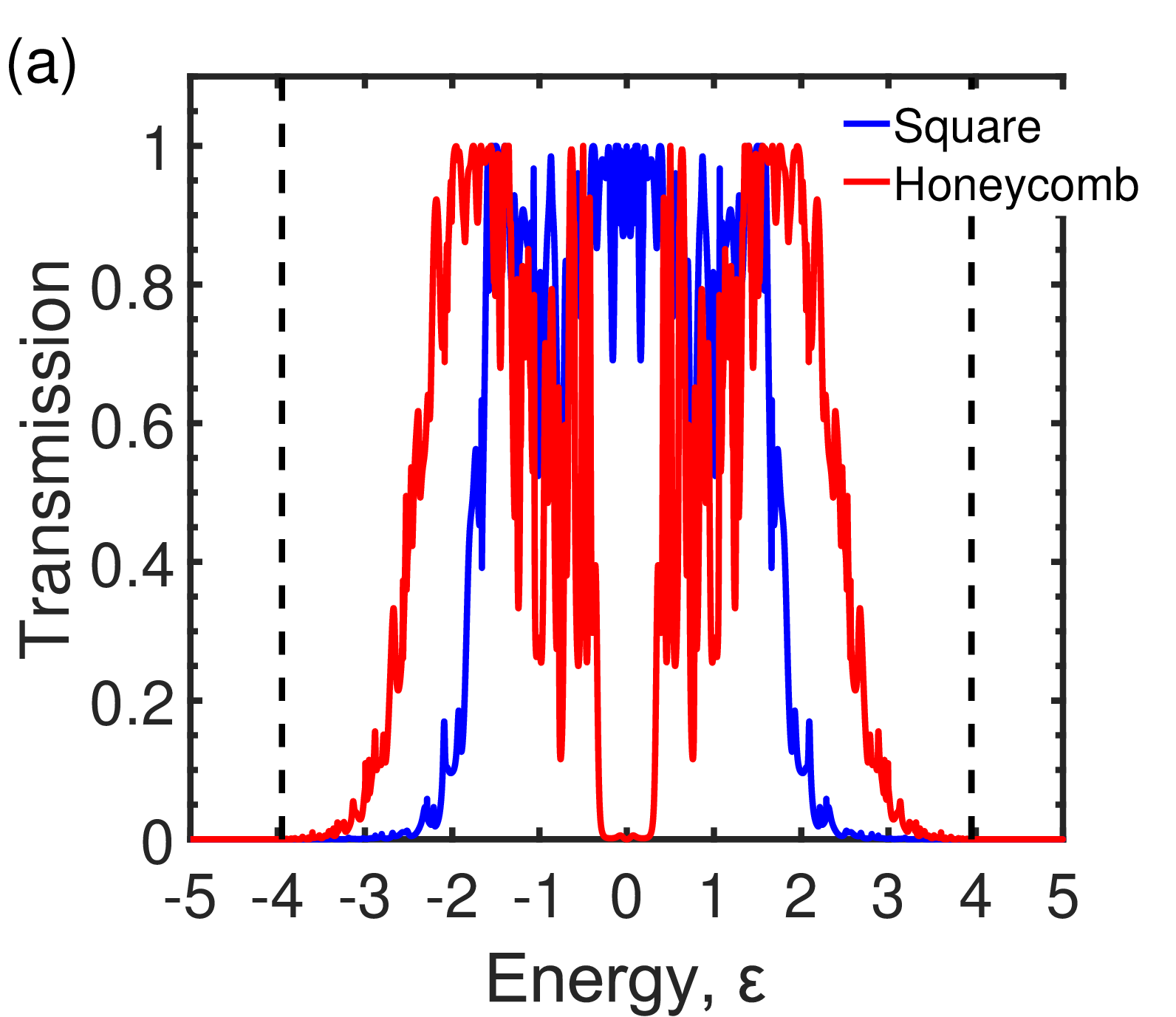}}~~
\subfloat{\label{fig:DOS}\includegraphics[width=0.24\textwidth]{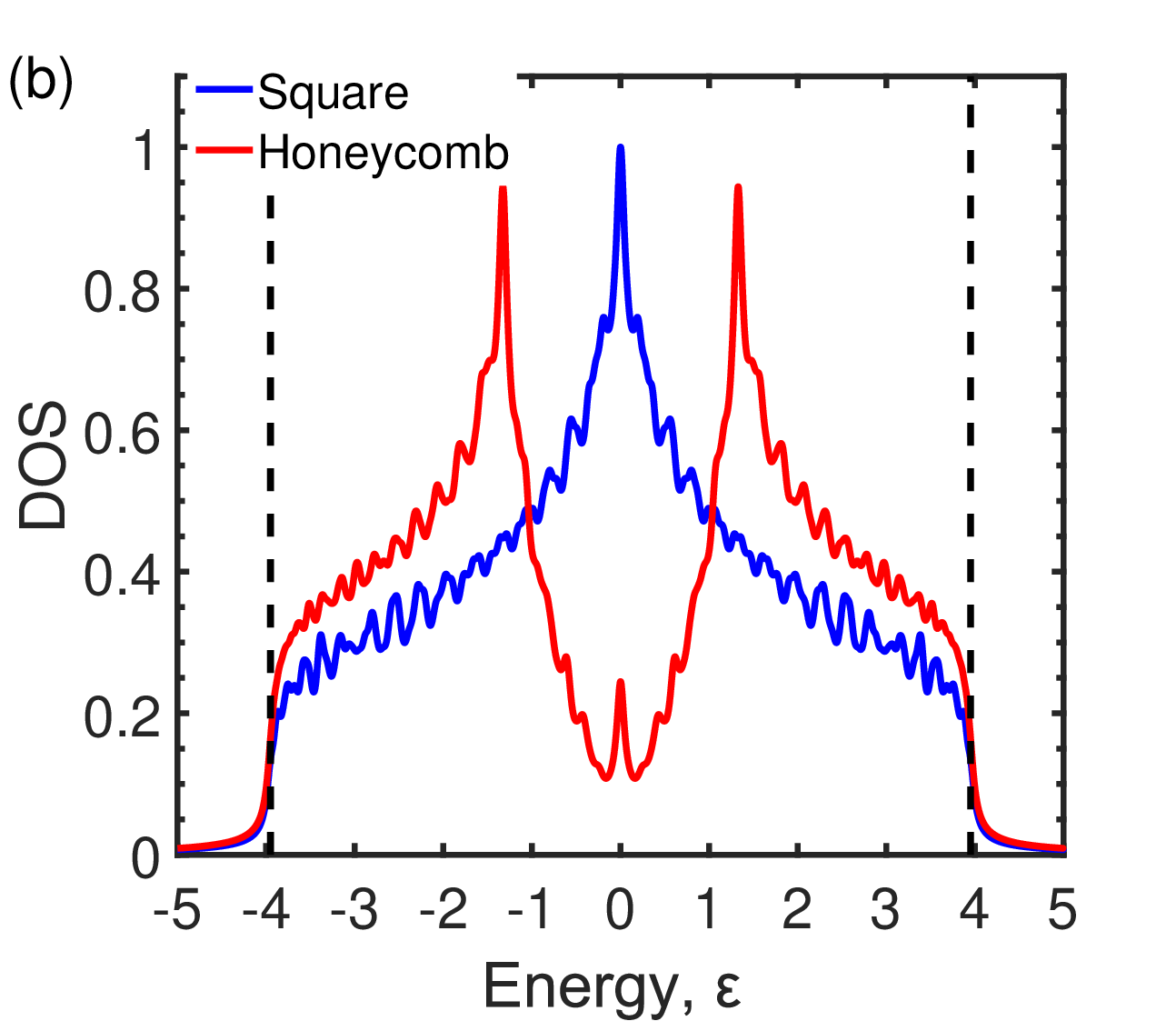}}~~
\caption{\label{fig:3t}(a) The transmission spectrum of two symmetric junctions. The black dotted lines indicate the bandwidth of the systems. (b) The density of states of the two symmetric junctions.}
\end{figure}

We can also analyze the transport coefficients of the junction in the grand canonical ensemble using the Landauer formalism. For this purpose, 
we introduce two fictitious baths coupling to the boundary sites of the two reservoirs, respectively. 
The linear transport coefficients $G$, $\alpha _c$ and $G_T$ of the junction are given by the following integrals

\begin{equation}
  K_n=\int_{\varepsilon _{\min}}^{\varepsilon _{\max}}{\frac{d\varepsilon}{2\pi}\left( \varepsilon -\mu \right) ^n\mT \left( \varepsilon \right) \left( -\frac{\partial f}{\partial \varepsilon} \right)}
  \label{equ:13}
\end{equation}
where $f=T/\left( \varepsilon-\mu \right)$ is the R-J distribution function and $\mT \left( \varepsilon \right)$ is the energy-dependent transmission coefficient, a central quantity in the Landauer formalism.
It is given by the Caroli formula from the nonequilibrium Green's function (NEGF) method as
\begin{equation} 
  \mT \left( \varepsilon \right) =\text{Tr}\left[ G_{d}^{\dag}\left( \varepsilon \right) \Gamma _L\left( \varepsilon \right) G_d\left( \varepsilon \right) \Gamma _R\left( \varepsilon \right) \right] 
  \label{equ:14}
\end{equation}
where $G_d=1/\left[ \left( E+i\eta \right) I-H-\Sigma _L-\Sigma _R \right] $ and $G_d^\dagger$ are the retarded and advanced Green's functions of the system, respectively.
$\eta$ is a small positive quantity,
$\Sigma _\alpha \left( \varepsilon \right)$ is the retarded self-energy due to coupling to bath $\alpha$, and $\Gamma _\alpha\left( \varepsilon \right) =-2\text{Im}\left[ \Sigma _\alpha\left( \varepsilon \right) \right]$ is the corresponding broadening function.
Under the wide band approximation, $\Sigma _\alpha$ can be regarded as an energy-independent constant, $\Sigma _\alpha=-i\gamma $, and $\gamma$ is a real parameter, which we choose as $\gamma = 0.1$.
The power and thermal current are written respectively as
\begin{align}
\label{eq:jp}
    J_P &= \int_0^{+\infty}d\varepsilon \mathcal{T}(\varepsilon) (f_L(\varepsilon)-f_R(\varepsilon)),\\
    J_Q &= \int_0^{+\infty}d\varepsilon (\varepsilon-\mu) \mathcal{T}(\varepsilon) (f_L(\varepsilon)-f_R(\varepsilon)).
\label{eq:jq}
\end{align}
The density of states (DOS) of the system is obtained by 
$\rho \left( \varepsilon \right) =i\text{Tr}\left( G_d-G_{d}^{\dag} \right) /M$.
Figure~\ref{fig:3t} shows the transmission spectra and DOS of the two symmetric junctions, from which the transport coefficients of the systems can be calculated and analyzed.

\end{document}